\documentclass[aps,prx,twocolumn,superscriptaddress,longbibliography]{revtex4-2}

\usepackage{graphics}
\usepackage{graphicx}
\usepackage{bbm,bm}
\usepackage{amsfonts}
\usepackage{amsmath}
\usepackage{xcolor}

\ifx\pdftexversion\undefined
\usepackage[dvips]{hyperref}
\else
\usepackage{hyperref}
\fi
\hypersetup{
  colorlinks = true, linkcolor = blue
}

\begin{document}
\preprint{}

\title{Full counting statistics of interacting lattice gases after an expansion: \\
The role of the condensate depletion in the many-body coherence
}

\author{Ga\'etan Herc\'e}
\email{These authors contributed equally to this work.}
\affiliation{Universit\'e Paris-Saclay, Institut d'Optique Graduate School, CNRS, Laboratoire Charles Fabry, 91127, Palaiseau, France}
\author{Jan-Philipp Bureik}
\email{These authors contributed equally to this work.}
\affiliation{Universit\'e Paris-Saclay, Institut d'Optique Graduate School, CNRS, Laboratoire Charles Fabry, 91127, Palaiseau, France}
\author{Antoine T\'enart}
\affiliation{Universit\'e Paris-Saclay, Institut d'Optique Graduate School, CNRS, Laboratoire Charles Fabry, 91127, Palaiseau, France}
\author{Alain Aspect}
\affiliation{Universit\'e Paris-Saclay, Institut d'Optique Graduate School, CNRS, Laboratoire Charles Fabry, 91127, Palaiseau, France}
\author{Alexandre Dareau}
\affiliation{Universit\'e Paris-Saclay, Institut d'Optique Graduate School, CNRS, Laboratoire Charles Fabry, 91127, Palaiseau, France}
\author{David Cl\'ement}
\affiliation{Universit\'e Paris-Saclay, Institut d'Optique Graduate School, CNRS, Laboratoire Charles Fabry, 91127, Palaiseau, France}

\date{\today}

\begin{abstract}
We study the full counting statistics (FCS) of quantum gases in samples of thousands of interacting bosons, detected atom-by-atom after a long free-fall expansion. In this far-field configuration, the FCS reveals the many-body coherence from which we characterize iconic states of interacting lattice bosons by deducing the normalized correlations $g^{(n)}(0)$ up to the order $n=6$. In Mott insulators, we find a thermal FCS characterized by perfectly-contrasted correlations $g^{(n)}(0)= n!$. In interacting Bose superfluids, we observe small deviations to the Poisson FCS and to the ideal values $g^{(n)}(0)=1$ expected for a pure condensate. To describe these deviations, we introduce a heuristic model that includes an incoherent contribution attributed to the depletion of the condensate. The predictions of the model agree quantitatively with our measurements over a large range of interaction strengths, that includes the regime where the condensate is strongly depleted by interactions. These results suggest that the condensate component exhibits a full coherence $g^{(n)}(0) =1$ at any order $n$ up to $n=6$ and at arbitrary interaction strengths. The approach demonstrated here is readily extendable to characterize a large variety of interacting quantum states and phase transitions.
\end{abstract}

\maketitle 

\section{Introduction}

The dispersion of a physical quantity contains important information, beyond that obtained from its average value.  The analysis of quantum and thermal noise is central in various systems, ranging from quantum electronics \cite{blanter2000} and quantum optics \cite{gardiner2004} to quantum gases \cite{burt1997, altman2004, schweigler2017}. The ultimate precision on the measurement of noise is given by the full counting statistics (FCS) \cite{levitov1996}, which is obtained with single-particle-resolved detection methods that provide the number of particles detected in a given time and/or space interval. These methods yield high-order moments of the particle number beyond the variance. Probing high-order moments is a means to study quantum phase transitions \cite{ivanov2010, gomez-ruiz2020, devillard2020}, universality \cite{eisler2013, lovas2017}, entanglement properties \cite{klich2009} or out-of-equilibrium dynamics \cite{esposito2009}. The FCS has indeed successfully characterized various phenomena in mesoscopic conductors \cite{levitov1996, blanter2000, gustavsson2006, maisi2014}, Rydberg \cite{liebisch2005, malossi2014, zeiher2016} and non-interacting  \cite{ottl2005, perrier2019} atomic gases. 

From a Quantum Information perspective, the FCS holds great promises for large ensembles of particles. In contrast to a full-state tomography \cite{flammia2012}, the FCS is accessible even in large systems as it probes information only about the diagonal part of the $n$-body density matrices ({\it i.e.} populations). Although it does not contain the total information about the quantum state, the FCS is sufficient to identify many quantum states without resorting to a consuming tomography. A similar idea was introduced by R. Glauber to characterize light sources from photon correlations at any order \cite{glauber1963}. For Gaussian states, for which the Wigner function is positive \cite{gardiner2004}, measuring the FCS or the magnitudes of correlation functions is indeed equivalent. 

In strongly-correlated quantum states characterized by non-Gaussian Wigner functions, measuring the FCS and many-body correlations is expected to reveal the non-trivial nature of such states \cite{schweigler2017, dolgirev2020, fabre2020, walschaers2021}. Moreover, recent works have shown that applying random unitary transformations before measuring the FCS provides access to non-diagonal correlators \cite{brunner2021, naldesi2022}, further motivating the development of experimental approaches to the FCS in strongly-interacting quantum systems.

\begin{figure*}[ht!]
\includegraphics[width=2 \columnwidth]{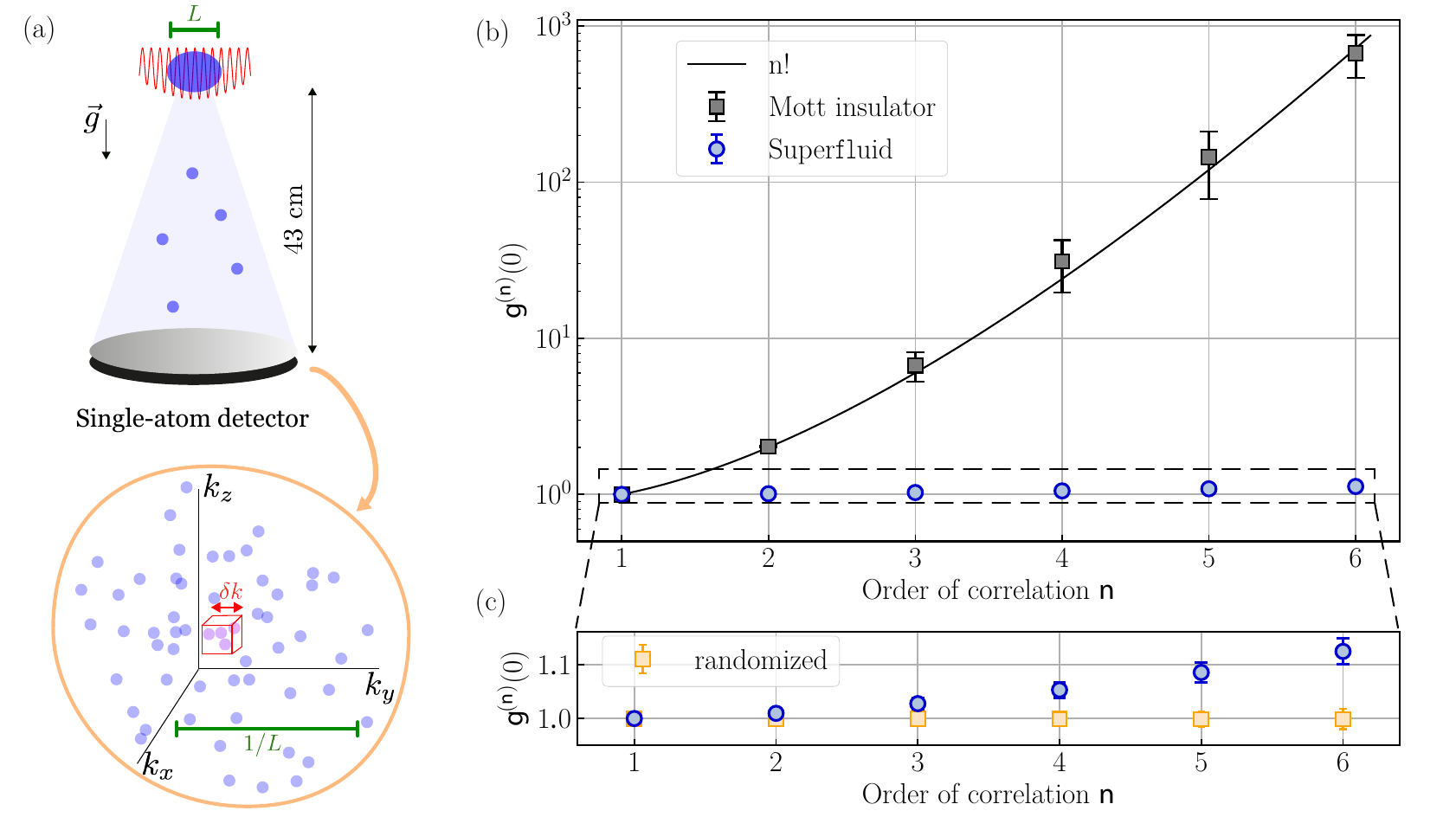}
\caption{
{\bf (a)} Free-fall expansion of interacting quantum gases of metastable Helium-4 atoms from a three-dimensional optical lattice, yielding the 3D positions of individual atoms in the momentum space. The full counting statistics $P(N_{\Omega})$ describes the statistics of the atom number $N_{\Omega}$ detected in a small voxel of volume $V_{\Omega} \sim (\delta k)^3$ (red cube). To reveal the many-body coherence properties of the trapped gas of size $L$, $\delta k$ is chosen such that $\delta k \ll 2\pi/L$. 
{\bf (b)} Magnitudes $g^{(n)}({\bm 0})$ of $n$-body correlations as a function of the order $n$, measured in a Mott insulator (black squares) and a superfluid (blue circles). The black solid line is the prediction for thermal states, $g^{(n)}(0)=n!$. 
{\bf (c)} $g^{(n)}({\bm 0})$ in the superfluid (blue circles) and in a randomized set (orange squares, see main text). A deviation to the prediction for a pure coherent state, $g^{(n)}(0)=1$, is observed.
}
\label{fig1}
\end{figure*}

In this letter, we report the measurement of the full counting statistics in large three-dimensional (3D) ensembles ($\sim 5 \times 10^3$ atoms) of interacting lattice bosons after a free-fall expansion (see Fig.~\ref{fig1}(a)). This configuration is analogous to the far-field regime of light propagation during which interferences take place, and after which the FCS identifies quantum states through their many-body coherence \cite{aspect2019}. In quantum gases, far-field -- or momentum -- correlations have been measured with single-atom detection in non-interacting and non-degenerate bosonic \cite{schellekens2005,dall2013} and fermionic \cite{jeltes2007, bergschneider2019} gases and in Bose-Einstein condensates \cite{hodgman2011}. More recently momentum correlations in interacting lattice bosons \cite{carcy2019, cayla2020, tenart2021} and interacting fermions \cite{holten2022} were studied. However, the FCS has thus far been measured only in 1D non-interacting bosons \cite{dall2013}. Here, we study various regimes of interacting 3D Bose gases across the superfluid-to-Mott phase transition, extending the measurement of many-body correlations to the strongly-interacting regime. This allows us to reveal the role of the depletion of the condensate in the many-body coherence properties of interacting Bose superfluids.

We characterize the FCS by measuring the probability distribution of the atom number $N_{\Omega}$ falling in a small voxel $V_{\Omega}$ (see Fig.~\ref{fig1}(a)). As explained below, a crucial asset of our work is the ability to probe many-body coherence into volumes smaller than that occupied by one mode in momentum space, {\it i.e.} $V_{\Omega} \ll (2 \pi/L)^3$ with $L$ the in-trap size of the gas. This possibility is given by the large quantum efficiency of our detector($\eta=0.53(2)$) \cite{tenart2021}.  Furthermore, we determine the magnitudes $g^{(n)}(0)$ of correlation functions (up to $n=6$) from the factorial moments of $N_{\Omega}$ \cite{mandel_wolf_1995}. As shown in Fig.~\ref{fig1}(b), $g^{(n)}(0)$ is found to vary by several orders of magnitude between the superfluid and the Mott insulator states. Interestingly, we observe small deviations to the predictions for a pure condensate  in Bose superfluids (see Fig.~\ref{fig1}(c)), which is in contrast to a previous work \cite{hodgman2011}. Since ab-initio calculations of many-body correlations with thousands of interacting atoms are beyond state-of-the-art numerics, we interpret our findings from introducing a heuristic model that includes the contribution of the condensate depletion. Its hypotheses rely on an intuitive physical picture in the weakly-interacting regime but, surprisingly, its predictions are in quantitative agreement with our observations even in the strongly-interacting regime.  Our observations lead us to conclude that the condensate component exhibits a full coherence, $g^{(n)}(0)=1$ (at least up to $n=6$), at arbitrary strength of interactions.

\section{Full Counting Statistics (FCS) of pure-state BECs and Mott insulators}\label{FCS-purestates}

Textbook descriptions of Bose-Einstein Condensates (BECs) and Mott insulators are based on pure states. Bose-Einstein condensation is associated with the breaking of phase symmetry \cite{pitaevskii-book} whose complex order parameter defines a coherent state describing the BEC. In a grand canonical approach, coherent states have a Poisson counting statistics, $P(N_{\Omega})=\langle N_{\Omega} \rangle^{N_{\Omega}}  \exp[-\langle N_{\Omega} \rangle] /N_{\Omega}!  $ where $N_{\Omega}$ is the number of detected bosons in the considered volume $V_{\Omega}$, and a full coherence  $g^{(n)}=1$ at any order $n$ of normalized correlations either in position or in momentum space  \cite{glauber1963}. In our experiment, we determine  
\begin{equation}
g^{(n)}(0)=g^{(n)}({\bm k},{\bm k},....,{\bm k})=\frac{\langle [a^{\dagger}({\bm k})]^n [a({\bm k})]^n \rangle }{\langle a^{\dagger}({\bm k}) a({\bm k}) \rangle ^n},
\end{equation}
where ${\bm k}$ is the momentum at which correlations are evaluated, {\it i.e.} where the volume $V_{\Omega}$ is located. For a pure coherent state we expect $g^{(n)}(0)=1$ at all orders. In contrast, a ``perfect'' Mott insulator -- a uniform Mott insulator at zero temperature --  is thought of as a Fock state in the (in-trap) position basis. In the momentum basis, which is probed after a long expansion from the trap, it is expected to exhibit thermal statistics \cite{folling2005, toth2008, carcy2019}. This is because far-field correlations reflect multi-particle interferences from a discrete series of emitters (atoms in the lattice sites) with no coherence between the sites (no tunnelling), a situation analog to that of light emitted by many incoherent sources. Thermal states are characterized by a counting statistics  $P(N_{\Omega})=(1-q)q^{N_{\Omega}}$ where $q=\langle N_{\Omega} \rangle/1+\langle N_{\Omega} \rangle$ and $g^{(n)}({\bm k},{\bm k},....,{\bm k})=n!$ \cite{milburn2008}. Note that both the Poisson and the thermal FCS are fully determined by a single parameter, the average number of particles $\langle N_{\Omega} \rangle$, as a result of the Gaussian character of their quantum state \cite{gardiner2004}. For Gaussian states, a detection efficiency $\eta$ smaller than one does not affect the measurement of the FCS -- nor that of $g^{(n)}$.

Whether these many-body coherence properties of pure states describe experiments is not granted. On the one hand, pure states are approximated descriptions of states produced in an experiment because of the coupling to the environment. Determining the level -- {\it e.g.} the order $n$ of correlations -- up to which a description in terms of pure states is valid provides a quantitative certification of experimental realizations. In the context of the development of platforms for quantum technologies, such a certification is of interest. On the other hand, the properties of quantum states produced at thermodynamical equilibrium are affected by constraints on macroscopic quantities. In the canonical (or micro-canonical) ensemble there are no fluctuations of the \emph{total} BEC atom number $N_{\rm BEC}$ in a gas of non-interacting bosons at zero temperature \cite{kristensen2019}: $N_{\rm BEC}$ is fixed to the total atom number, $N_{\rm BEC}=N$. The statistics of $N_{\rm BEC}$ is therefore not that of a coherent state \cite{castin1998}. To alleviate such global constraints and mimic a grand canonical ensemble, one may probe the FCS in a volume $V_{\Omega}$ much smaller than that, $V_{\rm BEC}$, occupied by the BEC. The volume $V_{\rm BEC}-V_{\Omega} \sim V_{\rm BEC}$ then serves as a reservoir for the sub-volume $V_{\Omega}$ where the number of bosons $N_{\Omega}\ll N$ may fluctuate \cite{NoteFluct}. 

Measuring the FCS in small sub-volumes $V_{\Omega}$ is also crucial if correlation functions $g^{(n)}({\bm k},{\bm k'},....,{\bm k''})$ are bell shaped with widths $l_{c}^{(n)}$. While this is not an issue for a coherent state, which is fully coherent over the entire volume it occupies, correlation functions of a thermal state must be probed in a volume $V_{\Omega}$ much smaller than the coherence volume $V_{c}^{(n)} =\large [ l_{c}^{(n)} \large ]^3$, since particles distant by $l_{c}^{(n)}$ are essentially uncorrelated. This is well known in the case of the Hanbury-Brown and Twiss effect where the property $g^{(2)}({\bm k},{\bm k'})=2$ is expected only if $|{\bm k}-{\bm k'}|$ is less that the width of the far-field diffraction pattern associated with the source size. 
In the far-field, the correlation lengths of Mott insulators and of thermal Bose gases are set by the inverse in-trap size, $l_{c}^{(n)} \sim 2 \pi/L$ \cite{carcy2019, cayla2020}. As a result, observing fully-contrasted $n$-body correlations requires using $V_{\Omega} \ll (2 \pi/L)^3$. This condition also ensures the above-mentioned criterion on probing BECs as the volume occupied by the BEC in the momentum space is set by $\Delta k \sim 1.6/L$ \cite{stenger1999}. With these considerations in mind, we compute the magnitudes $g^{(n)}(0)$ of correlation functions in a volume $V_{\Omega}\sim (\delta k)^3$ of the momentum space such that $\delta k \times L \ll 2 \pi$ (see Fig.~\ref{fig1}(a)). As illustrated below, this choice is essential to correctly reveal the full statistical properties.

\section{Measurement of the FCS in Mott insulators and Bose superfluids}

Our measurement of the FCS in the far-field exploits the 3D atom-by-atom detection of metastable Helium-4 ($^4$He$^*$) after a long free-fall expansion \cite{cayla2018, tenart2020}  (see Fig.~\ref{fig1}(a)). The detection efficiency is $\eta=0.53(2)$ per atom, with negligible dark counts. For an individual run, we register the number of atoms in each of the voxels $V_{\Omega}$ mentioned above, and we use more than 2000 runs to obtain the probability distribution in each voxel. We apply this approach to probe equilibrium quantum states of $^4$He$^*$ atoms loaded in the lowest energy-band of a three-dimensional (3D) optical lattice \cite{carcy2021}. The lattice implements the 3D Bose-Hubbard Hamiltonian whose main parameters are the tunnelling amplitude $J$ and the on-site (repulsive) interaction $U$. 

\begin{figure}[h!]
\includegraphics[width=\columnwidth]{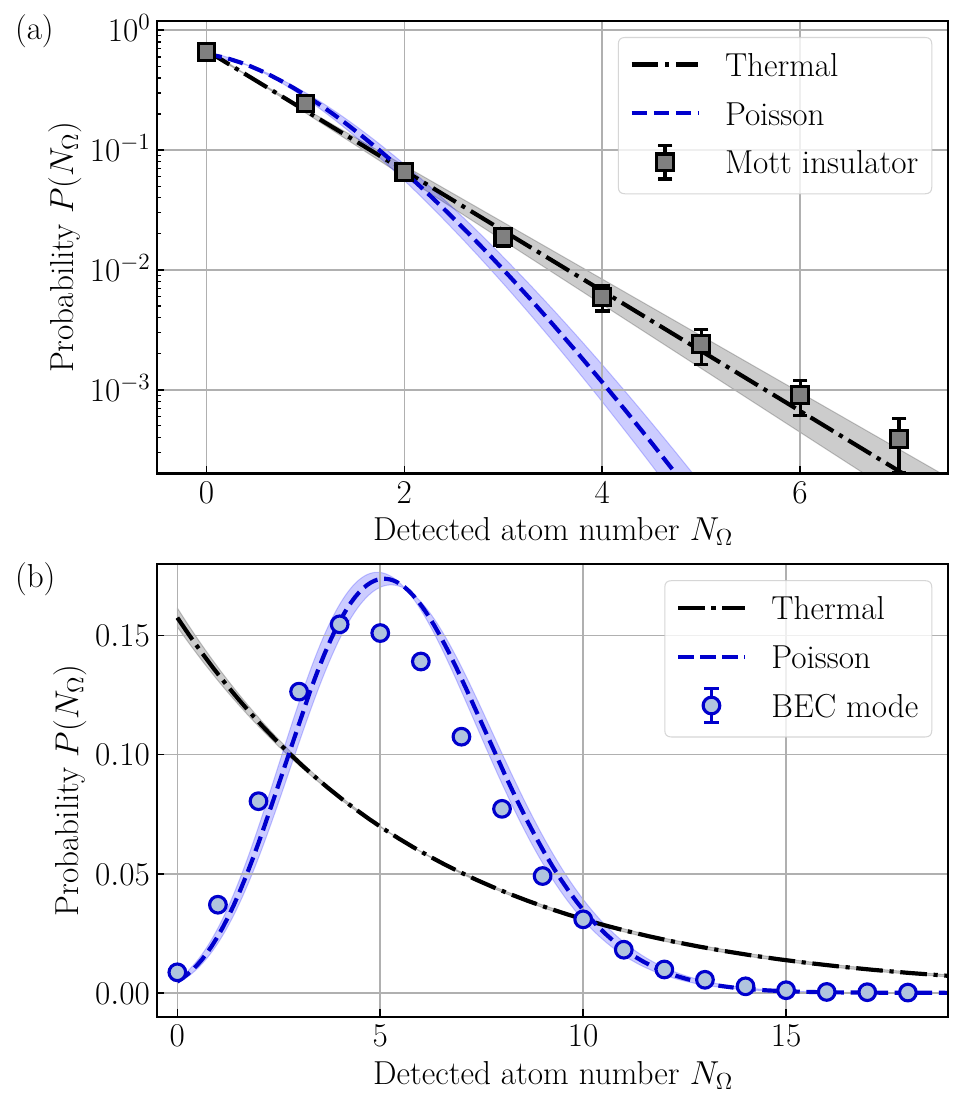}
\caption{
{\bf (a)} Full Counting statistics $P(N_{\Omega})$ to find $N_{\Omega}$ atoms in a small volume $V_{\Omega}$ of the momentum space when probing a Mott insulator with unity filling (black circles). The predictions for thermal ({\it resp.} Poissonian) statistics is shown as a dashed-dotted black ({\it resp.} dashed blue) line. We use the measured value $\langle N_{\Omega} \rangle = 0.46(5)$ for the theoretical predictions (the shaded areas reflect the uncertainty on $\langle N_{\Omega} \rangle$). The error bars denote the statistical uncertainty (standard deviation) estimated with the bootstrapping method. {\bf (b)} Same as (a) in the BEC mode (${\bf k}={\bf 0}$) of interacting lattice superfluids (SF) with $U/J=5$. The mean atom number is $\langle N_{\Omega} \rangle = 5.3(2)$. Error bars are smaller than the dots. 
}
\label{fig2}
\end{figure}

We first investigate the thermal nature of Mott insulators in the momentum space. We realize Mott insulators with $N=6.5(6)\times 10^3$ atoms at $U/J=76$, which corresponds to a lattice filling of one atom per site at the trap center \cite{carcy2019} and an almost uniform filling of the first Brillouin zone in the momentum space. To compute the counting statistics in this first set of experiments, we divide the first Brillouin zone into cubic voxels $V_{\Omega}$ of size $\delta k=6 \times 10^{-2} k_{d}$ and average the probability distributions measured over all these voxels. Here  $k_{d}=2 \pi/d$ is the momentum associated with the lattice spacing $d=775~$nm and the size $\delta k$ is comparable to that of one mode in momentum space, $\delta k \sim 2 \pi/L$ (see \ref{AppA}). The resulting FCS of the Mott state is shown in Fig.~\ref{fig2}(a). It is found to be in excellent agreement with a thermal statistics whose average atom number is that measured in the experiment, $\langle N_{\Omega} \rangle = 0.46(5)$. 

The statistical properties of the Mott state are also revealed through the magnitudes $g^{(n)}(0)$ of the normalized correlation functions, as an alternative to the probability distribution $P(N_{\Omega})$. We determine $g^{(n)}(0)$ from the factorial moments of the detected atom number $N_{\Omega}$ in small voxels $V_{\Omega}$ with $\delta k \ll 2 \pi/L$ (see \ref{AppB}),
\begin{equation}
g^{(n)}(0)=\frac{\langle N_{\Omega}(N_{\Omega}-1)...(N_{\Omega}-n+1)\rangle}{\langle N_{\Omega}\rangle^n},
\end{equation}
transposing a well-known approach in quantum optics \cite{laiho2022}. The correlation functions of quantum optics \cite{glauber1963} are defined with normal ordering of the destruction operators since a detected photon is destroyed. 
Even if it is not always emphasized, the correlation functions in quantum optics are indeed calculated from the factorial moments, and it is remarkable that the corresponding quantities in classical statistics are also based on factorial moments \cite{goodmanbook}. In the experiments reported here, each detected He$^*$ atom is destroyed ({\it i.e.} it decays to its ground-state) and the results of quantum optics are directly transposed.
As explained in section~\ref{FCS-purestates}, fully-contrasted correlation functions are measured only when computed in voxels of small size $\delta k \ll 2 \pi/L$. This requirement is more stringent than the one for measuring the probability distributions (see \ref{AppB}).

The results are plotted in Fig.~\ref{fig1}(b) and are found in excellent quantitative agreement with the prediction for thermal states, $g^{(n)}(0)=n!$. They represent a significant progress with respect to the literature where 2- \cite{folling2005} and 3-body  \cite{carcy2019} correlations had only been measured with limited amplitudes $g^{(n)}(0) < n!$. The momentum-space FCS of a Mott insulator is thus identical to that of a statistical mixture of thermal bosons. Their full correlation functions however differ in their sizes $l_{c}^{(n)}$ which are determined by their different in-trap sizes and the incompressible ({\it resp.} compressible) character of Mott insulators ({\it resp.} thermal gases) \cite{carcy2019}.

\begin{figure*}[ht!]
\includegraphics[width=2\columnwidth]{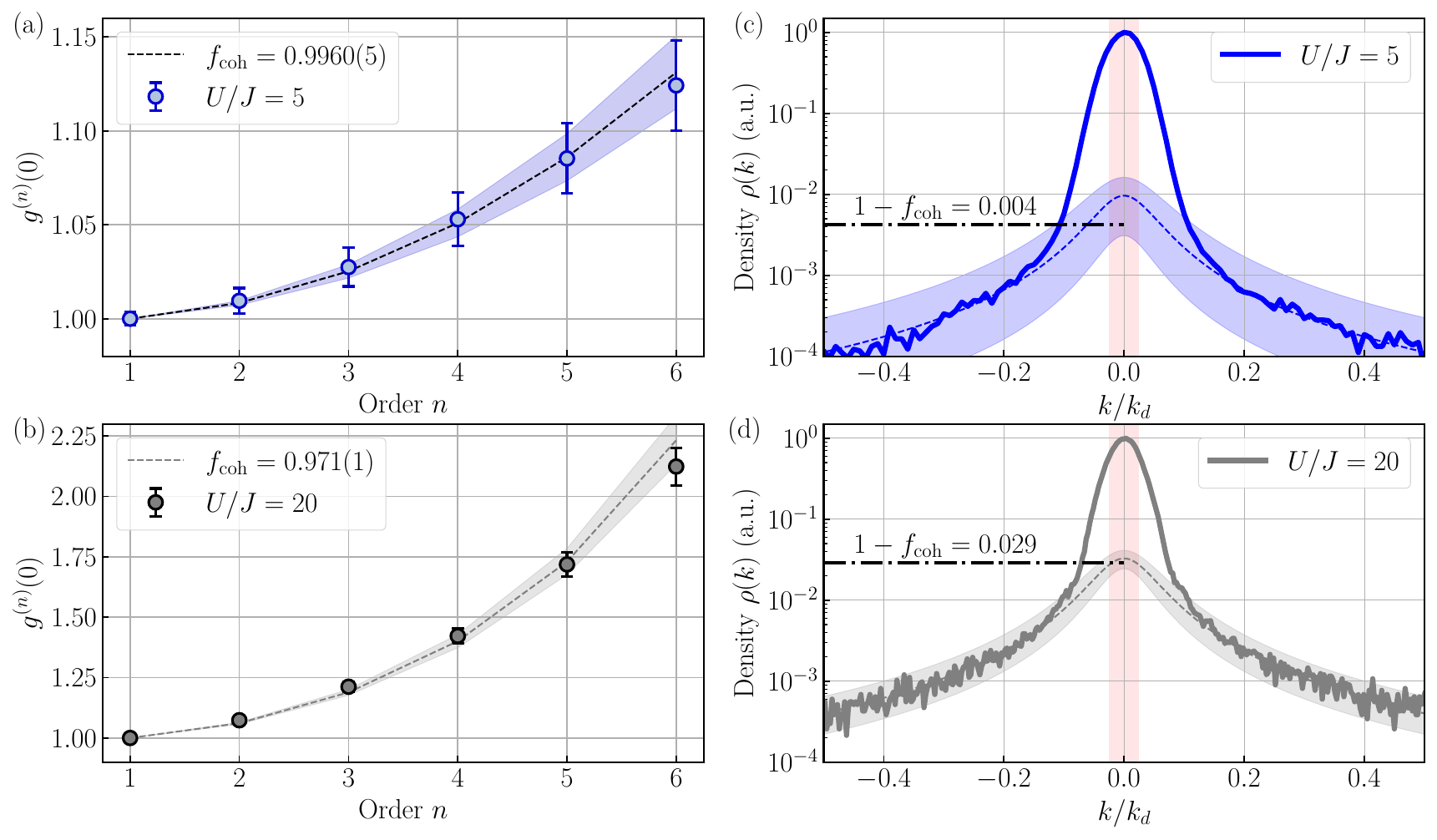}
\caption{
{\bf (a)}-{\bf (b)} Plots of $g^{(n)}(0)$ measured at ${\bm k}={\bm 0}$ in lattice superfluids with $U/J=5$ and $U/J=20$. The data in panel (a) is identical to that of shown in Fig.~\ref{fig1}(c). The dashed lines ({\it resp.} the shaded areas) are the predictions of the model with the values $f_{\rm coh}$ ({\it resp.} the uncertainties on $f_{\rm coh}$) fitted to the data. {\bf (c)}-{\bf (d)} Plots of 1D cuts through the momentum densities measured at $U/J=5$ and $U/J=20$ and normalized to their value at $k=0$. The vertical shaded area indicates the volume occupied by the sphere $S_{\Omega}$ where the counting statistics is evaluated. The horizontal dashed-dotted lines indicate the fitted values $1-f_{\rm coh}$. The dashed lines are Lorentzian functions fitted to the tails of the densities in the range $[0.2 k_{d}, 0.5k_{d}]$ in order to estimate the density of the depletion at $k=0$ (the shaded areas represent the error from the fitted parameters). 
}
\label{fig3}
\end{figure*}

In a second set of experiments, we address the $n$-body coherence of the BEC component in lattice superfluids with $N=5(1)\times 10^3$ at $U/J=5$. In the momentum space, the BEC occupies a volume of width $\Delta k  \simeq 0.15 k_{d}$ centered at ${\bm k}={\bm 0}$ \cite{tenart2021}. This volume contains many atoms in each run and the statistics of the atom number falling in a sphere $S_{\Omega}$  of a radius $\delta k=0.025 k_{d} \ll  \Delta k $, centered at ${\bm k}={\bm 0}$, is sufficient to extract the counting statistics (see \ref{AppC}). We measure the counting statistics $P(N_{\Omega})$ in $S_{\Omega}$ over about $2000$ experimental runs, with the results plotted in Fig.~\ref{fig2}(b). It is compared with Poisson and thermal statistics whose average atom number $\langle N_{\Omega}\rangle=5.3(2)$ is that measured in the experiment. The counting statistics in the BEC mode is close to the Poisson FCS and clearly differs from the thermal FCS. This is confirmed by the measured values of $g^{(n)}(0)$ calculated from the normalized factorial moments of $N_{\Omega}$. 
In Fig.~\ref{fig1}(b) we plot $g^{(n)}(0)$ as a function of $n$ and find that $g^{(n)}(0) \sim 1$ at any order $n$ in the BEC mode. These results, predicted by Glauber for a fully coherent sample, are in striking contrast with those of the Mott state, a difference that illustrates the outstanding capabilities of the FCS measured after an expansion to reveal the $n$-body coherence. 

Surprisingly, however, our measurements in the BEC mode deviate from the predictions for a coherent state and from a previous observation \cite{hodgman2011}: the deviation in the FCS (see Fig.~\ref{fig2}(b)) is reflected in the fact that $g^{(n)}(0) > 1$, as shown in Fig.~\ref{fig1}(c). To verify that the observed deviations are statistically meaningful, we apply our computation of the $n$-body correlations to a randomized set, with the same numbers of atoms and of runs. This randomized set is obtained by randomly shuffling the detected atoms across the experimental runs. Doing so, atom correlations present within individual runs ({\it i.e.} before shuffling) should vanish, and a Poisson statistics should be observed as a result of the discrete nature of our detection method applied to fully independent events. Indeed we find $g^{(n)}(0)=1.00(2)$ at any order $n$ of the randomized ensemble (see the orange squares in Fig.~\ref{fig1}(c)), confirming that the deviations in the (non-randomized) experimental data are significant. 
The randomization method also yields a Poisson statistics when applied to the Mott insulator data set. Note that the results of the randomization method validate the algorithm used to compute the $n$-body correlations and provide a means to test the accuracy of the measured statistics (see \ref{AppD}). In the next section, we propose an interpretation of the deviations $g^{(n)}(0) > 1$ revealed by our experiment in the case of lattice superfluids.  

\section{Coherent fraction of Bose superfluids in the BEC mode}

At thermal equilibrium, a fraction of the total atom number is expelled from the BEC by the finite interactions (quantum depletion) and by the finite temperature (thermal depletion). Even if it amounts to a negligible fraction of the atom number $N_{\Omega}$ falling in $S_{\Omega}$,  the total depletion of the condensate may contribute to the measured statistics. We define the ``coherent fraction" $f_{\rm coh}$ as the fraction of $N_{\Omega}$ that belongs to the BEC. Building on these considerations, we introduce a heuristic model that assumes {\it (i)} that atoms in the BEC and in the depletion contribute independently to the measured counting statistics in $S_{\Omega}$  \cite{castin1998}, and {\it (ii)} that the BEC is a coherent state while both the thermal and quantum depletion exhibit thermal statistics in $S_{\Omega}$. While a thermal FCS is expected for the thermal depletion, we emphasize that describing the contribution of the quantum depletion with a thermal statistics is an assumption. The statistics of the quantum depletion was shown to be thermal when measured at {\it non-zero} momenta, outside the BEC \cite{cayla2020}. Whether this is also valid for small momenta within the BEC mode is difficult to assess in a harmonic trap. 
Finally, there is \emph{a priori} no reason for our model to be valid in the strongly-interacting regime. In contrast to the weakly-interacting regime where the Bogoliubov approximation holds, neglecting the correlations between the condensate and its depletion at large interaction strengths could be an erroneous assumption. Our measurements however show that the model agrees with the experimental data over a wide range of interaction strengths.

With the hypotheses of our model, we obtain an analytical prediction for $g^{(n)}(0)$ that depends only on the coherent fraction $f_{\rm coh}$ (see  \ref{AppE}),
\begin{equation}
g^{(n)}(0) - 1 = \sum_{p=1}^{n-1} \left [ (n-p)!  \binom{n}{p}^2 - \binom{n}{p} \right ] f_{\rm coh}^p (1-f_{\rm coh})^{n-p}  
\label{Eq:gn}
\end{equation}
While our model straightforwardly predicts the magnitudes $g^{(n)}(0)$, this is not the case for the probability distribution $P(N_{\Omega})$. The latter results from a complex convolution of the probability distributions of the condensate and of the depletion and it is well established that obtaining $P(X)$ from the moments of  a random variable $X$ is a difficult problem \cite{akhiezerbook}. 

In Fig.~\ref{fig3}(a), we fit the experimental data shown in Fig.~\ref{fig1}(b) with the analytical prediction of Eq.~(\ref{Eq:gn}), finding a good agreement with the coherent fraction as the only adjustable parameter (found equal to $f_{c}=0.9960(5)$ for the case of $U/J =5$). We then repeated our measurements for various  condensate fractions. To do so, we change the lattice depth to obtain ratios of on-site interaction $U$ to tunnelling amplitude $J$ ranging from $U/J=2$ to $U/J=22$. In this range of parameters, the gas remains far from entering the Mott insulator regime expected at the critical ratio $(U/J)_{c}\simeq 25 - 30$ \cite{carcy2021}, but it enters the strongly-interacting regime where the condensate is strongly depleted (at $U/J=22$ the condensate fraction is $f_{c}\sim0.15$). 

In Fig.~\ref{fig3}(b) we plot the magnitude of the $n$-body correlations for a lattice superfluids with an increased interaction $U/J=20$. The deviation from the ideal coherent state is increased, in qualitative agreement with  the physical picture at the root of the model. To be quantitative, we fit the experimental data with the analytical prediction of Eq.~(\ref{Eq:gn}). Firstly, we confirm that Eq.~(\ref{Eq:gn}) correctly fits the values of $g^{(n)}(0)$ with a single adjustable parameter $f_{\rm coh}$. Secondly, the extracted values of $f_{\rm coh}$ decrease with the interaction strength as intuitively expected. The uncertainty on the values $f_{\rm coh}$ is extremely small, at the $\sim0.1\%$ level. As can be inferred from Fig.~\ref{fig3}, the larger the order $n$ of correlations we measure, the smaller the uncertainty on $f_{\rm coh}$. This illustrates the extreme sensitivity of high-order correlations to probe many-body coherence.

A quantitative test of the model would compare the value $1-f_{\rm coh}$ to the fraction $\eta_{D}$ of depleted atoms detected within $S_{\Omega}$. We are not aware of a quantitative analytical prediction for $\eta_{D}$ in 3D interacting trapped lattice Bose gases. However, an indirect comparison is amenable from measuring the momentum densities. In Fig~\ref{fig3}(c)-(d), we plot 1D cuts through the momentum densities measured at $U/J=5$ and $U/J=20$ and we fit the tails (in the range $[0.2k_{d},0.5k_{d}]$) with a Lorentzian function to extrapolate the density of the depletion at $k=0$. Using a Lorentzian function is an arbitrary choice which happens to correctly fit the tails. This analysis indicates that the values $1-f_{\rm coh}$ are compatible with the extrapolated densities, while both quantities vary by one order of magnitude. 

\begin{figure}[t!]
\includegraphics[width=\columnwidth]{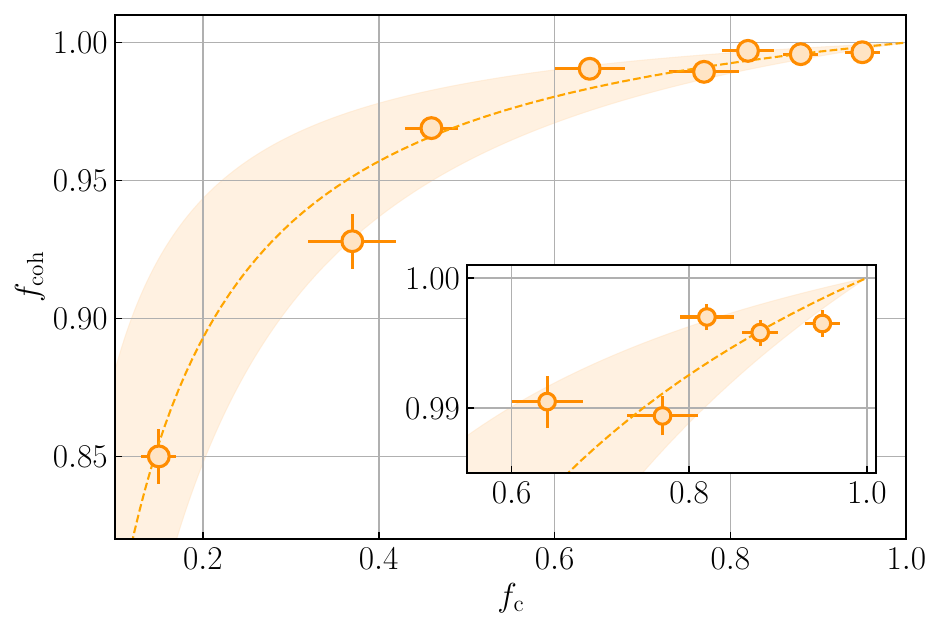}
\caption{
Coherent fraction $f_{\rm coh}$ as a function of the condensate fraction $f_{c}$. The dashed-line is the prediction of Eq.~(\ref{Eq:f_coh}) where the ratio $\mathcal{V}_{c}/\mathcal{V}_{d}=0.03(1)$ is evaluated from the measured density profiles (see \ref{AppF}). The shaded area reflects the uncertainty on $\mathcal{V}_{c}/\mathcal{V}_{d}$.
}
\label{fig4}
\end{figure}

To assess the validity of the model, we proceed with a quantitative comparison of the measured coherent fraction to the measured condensate fraction \cite{cayla2018}. The coherent fraction $f_{\rm coh}$ in $S_{\Omega}$ differs from the condensate fraction $f_{c}$ of the entire gas since the radius of $S_{\Omega}$ is much smaller than $2 \pi/L$. Moreover, the scaling of $f_{\rm coh}$ with $f_{c}$ is non-linear since $f_{\rm coh}$ reveals the BEC atom number in a small volume $S_{\Omega}$ where the BEC contribution is maximum while that of the depletion is small (except when $f_{c}\ll1$). The volumes occupied by the BEC ($\mathcal{V}_{c}$) and by the depletion ($\mathcal{V}_{d}$) set their respective contributions in $S_{\Omega}$. A simple estimate (see \ref{AppF}) leads to 
\begin{equation}
f_{\rm coh} \simeq \frac{f_{c}}{f_{c}+(1-f_{c})\mathcal{V}_{c}/\mathcal{V}_{d}}.
\label{Eq:f_coh}
\end{equation}

In Fig.~\ref{fig4} we plot the measured values of $f_{\rm coh}$ and $f_{c}$, along with the prediction of Eq.~(\ref{Eq:f_coh}). Here, $f_{c}$ is obtained similarly to \cite{cayla2018} and  the ratio $\mathcal{V}_{c}/\mathcal{V}_{d}$ used to plot Eq.~(\ref{Eq:f_coh}) is estimated from the measured density profiles (see  \ref{AppF}). The quantitative prediction of our model (without any adjustable parameter) correctly reproduces the observed non-linear dependency of $f_{\rm coh}$ with $f_{c}$. As discussed previously, the agreement in the strongly-interacting regime is rather surprising. Understanding this fact requires a more elaborate theoretical approach than the one presented in this work.

We observe that the heuristic model quantitatively captures the deviations to $g^{(n)}(0)=1$, attributing the latter to the presence of depleted atoms (both from the thermal and the quantum depletion). In turn, this suggests that BEC atoms have the statistics of a fully coherent state, with $g^{(n)}(0)=1$ at any order $n\leq6$. 
The role of the depletion in the many-body coherence unveiled in this work was not identified previously \cite{hodgman2011}. A major difference of our experiment with respect to that of \cite{hodgman2011} is that the volume $V_{\Omega}$ used in that work to compute the statistics is larger than the coherence volume $(2 \pi / L)^3$, a choice made in light of a smaller detection efficiency $\eta \sim 0.08$. Using too large of a volume $V_{\Omega}$ entails a convolution of the correlation functions that significantly reduces their amplitudes and, in turn, hides the contribution from the depletion. This is illustrated by the fact that in the thermal case the authors of  \cite{hodgman2011} find $g^{(2)}(0)=1.022(2)$ and $g^{(3)}(0)=1.061(6)$ instead of $g^{(2)}(0)=2$ and $g^{(3)}(0)=6$ . We conclude that our ability to obtain sufficient statistics in tiny volumes is a major asset to quantitatively probe many-body correlations.

\section{Conclusion}

We have presented measurements of the full counting statistics (FCS) and of perfectly-contrasted $n$-body correlations in interacting lattice Bose gases. We find that a Mott insulator exhibits a thermal statistics in the momentum space with $g^{(n)}(0)=n!$, while the condensate component is fully coherent with $g^{(n)}(0)=1$ (at least) up to $n=6$. The latter conclusion derives from assuming that the heuristic model we introduce to analyse the FCS in Bose superfluids and to describe the contribution from the depletion of the condensate is valid. We have assessed this validity in the experiment from studying the coherent fraction at increasing interaction strengths. A theoretical validation is beyond the scope of this work. Our results represent the most stringent certifications of the many-body coherence properties of Mott insulators and of BECs, and they validate emblematic pure-state descriptions of $n$-body coherence up to the order $n=6$.

In the future, the experimental approach we use is readily extendable to probe the many-body coherence in a large variety of interacting quantum states and phase transitions realized on cold-atom platforms ({\it e.g.} \cite{lovas2017, eisler2013, devillard2020, contessi2022}). It relies on {\it (i)} a free-fall expansion from the trap and {\it (ii)} a single-atom-resolved detection method. The first condition is ensured with lattice or low-dimensional (1D and 2D) gases for which interactions do not affect the expansion, as well as with quantum gases for which interactions are switched off during the expansion \cite{bloch2008}. The second condition is met with various detection methods and atomic species \cite{ott2016}. 

Another interesting direction consists in realizing recent proposals to access non-trivial $n$-body correlations \cite{brunner2021, naldesi2022}. These  theoretical works have shown that non-diagonal correlators can be accessed from combining two-particle unitary transformations -- {\it e.g.} beamsplitters -- with a measurement of the FCS. Such protocols have many potential applications, from quantifying entanglement to implementing variational algorithms.

\section*{Acknowledgement}

We thank S. Butera, A. Browaeys, I. Carusotto, T. Lahaye, T. Roscilde and the members of the Quantum Gas group at Institut d'Optique for insightful discussions. We acknowledge financial support from the R\'egion Ile-de-France in the framework of the DIM SIRTEQ, the ``Fondation d'entreprise iXcore pour la Recherche", the Agence Nationale pour la Recherche (Grant number ANR-17-CE30-0020-01). D.C. acknowledges support from the Institut Universitaire de France. A. A. acknowledges support from the Simons Foundation and from Nokia-Bell labs.


\bibliography{Herce2022.bib}

\begin{thebibliography}{55}
\bibitem [{Not()}]{NoteFluct}%
  \BibitemOpen
  \bibinfo {note} {Note that in the experiment the total atom number $N$ is not
  strictly constant from shot to shot. However, the measurement of the FCS in a
  tiny volume $V_{\Omega}$ is largely immune to small shot-to-shot fluctuations
  of $N$}\BibitemShut {NoStop}%
\end{thebibliography}


\setcounter{section}{0}
\renewcommand\thesection{Appendix \Alph{section}} 

\section{Probability distribution in Mott insulators and multimode thermal statistics}\label{AppA}

We consider the thermal statistics found in the case of Mott insulators. The bell-shaped correlation functions are well-fitted by a Gaussian function \cite{carcy2019} and we introduce the two-body correlation length $l_{c}$ as
\begin{equation}
g^{(2)}({\bm k}_{1},{\bm k}_{2})=g^{(2)}(0) \times \exp \left ( \frac{- 2 |{\bm k}_{1} - {\bm k}_{2} |^2}{l_c^2} \right ).
\end{equation}
For the data set of a Mott insulator shown in the main text, the two-body correlation length is $l_{c}/k_{d}= 0.031(2)$ and corresponds to the expected value \cite{carcy2019}, $l_{c}/k_{d}\sim 1/N_{\rm site}$ where $N_{\rm site}$ is the number of lattice sites occupied by the Mott insulator in the trap. In the case of a thermal statistics, the two-body correlation length in momentum space is equal to the inverse trap size of the gas, {\it i.e.} to the size of a mode in momentum space. We define a mode from its coherence volume $V_{c}$  as a cubic voxel of size $2 l_{c}$. 

When one computes two-body correlations in a cubic volume $V_{\Omega}$ of size $\delta k$ comparable to, or larger than, the correlation length $l_{c}$, the measured magnitude differs from the fully-contrasted value $g^{(2)}(0)$ due to the spatial averaging of $g^{(2)}({\bm k}_{1},{\bm k}_{2})$. For instance, using $V_{\Omega}=V_{c}$, the magnitude of two-body correlations is reduced by a factor $1/( \sqrt{\pi/8} \ {\rm erf}[\sqrt{2}] )^3 \sim 4.7$. In contrast, we find that computing the probability distribution $P(N_{\Omega})$ is not affected by considering a volume $V_{\Omega} \sim V_{c}$, as we shall now discuss.
\\

\begin{figure}[h!]
\includegraphics[width=\columnwidth]{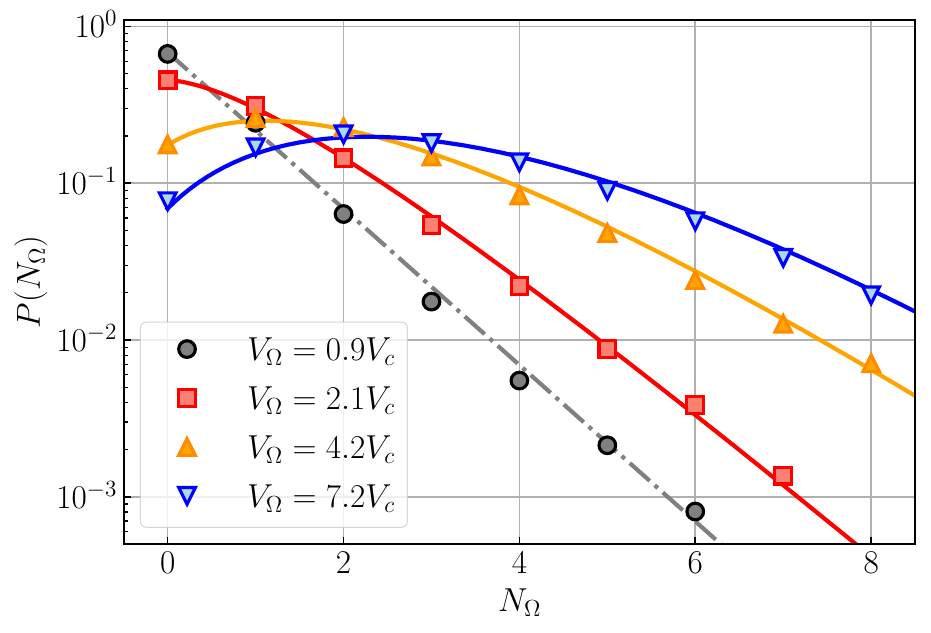}
\caption{
Probability distributions $P(N_{\Omega})$ measured in a Mott insulator with a varying voxel size $V_{\Omega}$. For $V_{\Omega}= 0.9 V_{c}$, the measured distribution (black dots) matches the prediction (black dashed-dotted line) for a thermal distribution. For $V_{\Omega}>V_{c}$, the measured distributions (red squares, orange triangles and blue triangles) agree with the predictions $P_{M}(N_{\Omega})$ for a multi-mode thermal distribution (red,  orange and blue solid lines). No adjustable parameters are used to plot the predictions $P_{M}(N_{\Omega})$ that depend on the measured mean atom number $\langle N_{\Omega} \rangle$ and the chosen ratio $M=V_{\Omega}/V_{c}$.
}
\label{fig-supp0}
\end{figure}

In Fig.~\ref{fig-supp0} we plot the probability distributions $P(N_{\Omega})$ computed with volumes $V_{\Omega}$ equal or larger than $V_{c}$. Firstly, we observe that the statistics is thermal when  $V_{\Omega} \leq V_{c}$. More specifically, there is no need to ensure the condition $V_{\Omega} \ll V_{c}$ to measure the probability distribution of a thermal statistics. Secondly, this statement is confirmed by analysing the measured probability distributions $P(N_{\Omega})$ for $V_{\Omega} > V_{c}$. Indeed, the probability distribution computed over a number $M$ of independent modes, each of which exhibiting a thermal statistics with an average occupation $\langle N \rangle$, reads \cite{goodmanbook, perrier2019}
\begin{equation}
P_{M}(N_{\Omega}) = \frac{(\langle N_{\Omega} \rangle+M-1)!}{\langle N_{\Omega} \rangle!(M-1)!} \frac{(\langle N_{\Omega} \rangle /M)^{N_{\Omega}}}{(1+\langle N \rangle/M)^{N_{\Omega}+M}}.
\end{equation}
In Fig.~\ref{fig-supp0} the prediction $P_{M}(N_{\Omega})$ for a multi-mode thermal statistics is plotted with a number of modes equal to $M=V_{\Omega}/V_{c}$. These predictions are found to be in excellent agreement with the measured distributions without any adjustable parameters. 

\section{Magnitudes of correlation functions in Mott insulators}\label{AppB}

\subsection{Condition to measure fully-contrasted atom correlations}

As discussed in section~\ref{FCS-purestates}, obtaining fully-contrasted $n$-body correlation functions requires to probe the statistics in small voxels $V_{\Omega}\ll V_{c}$. This condition is much stringent than that ($V_{\Omega}\leq V_{c}$) to obtain the probability distribution $P(N_{\Omega})$. This poses serious difficulties in the case of a Mott insulator as its momentum-space density is  low because a Mott insulator occupies a large volume in momentum-space, extending beyond the first Brillouin zone. The statistics of the atom number $N_{\Omega}$ in a single voxel $V_{\Omega} \ll V_{c}$ is therefore not sufficient to extract the magnitude $g^{(n)}(0)$ of the $n$-body correlation functions accurately, even when using an average over all the voxels contained in the first Brillouin zone. To circumvent this issue, we first compute the statistics in larger voxels from which the results in small voxels $V_{\Omega} \ll V_{c}$ are then extrapolated. This procedure is detailed below. 

\subsection{Magnitudes of  correlation functions measured in voxels $V_{\Omega} \sim V_{c}$}

We first evaluate the magnitudes of correlation functions in anisotropic voxels of volume $V(\delta k_{\perp})= \delta k \times \delta k_{\perp}^2$ with $\delta k_{\perp} \geq l_{c} > \delta k = 0.015 k_{d}$. The volume $V(\delta k_{\perp}) \sim V_{c} $ is sufficiently large to compute the statistics of $N_{\delta k_{\perp}}$. 

To proceed, we divide the first Brillouin zone in voxels of volume $V(\delta k_{\perp})$ and label each voxel with an integer $j$. The factorial moment in one voxel is denoted $\langle N_{\delta k_{\perp}}(N_{\delta k_{\perp}}-1)...(N_{\delta k_{\perp}}-n+1)\rangle_{j}$ and the average of the factorial moments over the first Brillouin zone reads
\begin{equation}
\sum_{j} \langle N_{\delta k_{\perp}}(N_{\delta k_{\perp}}-1)...(N_{\delta k_{\perp}}-n+1)\rangle_{j}.
\end{equation} 
We obtain the magnitudes of correlation functions from normalizing the average value of the factorial moments, 
\begin{equation}
g^{(n)}_{\delta k_{\perp}}(0)=\frac{ \sum_{j} \langle N_{\delta k_{\perp}}(N_{\delta k_{\perp}}-1)...(N_{\delta k_{\perp}}-n+1)\rangle_{j}}{\sum_{ j} \langle N_{\delta k_{\perp}}\rangle_{j}^n}.
\end{equation}

In the case of two-body correlations, an alternative to the approach based on the factorial moments is to compute the histogram of atom pairs \cite{carcy2019}. We have verified that both approaches yield similar results for the magnitude $g^{(2)}_{\delta k_{\perp}}(0)$.

\subsection{Magnitudes of correlation functions measured in voxels $V_{\Omega} \ll V_{c}$}

\begin{figure}[h!]
\includegraphics[width=\columnwidth]{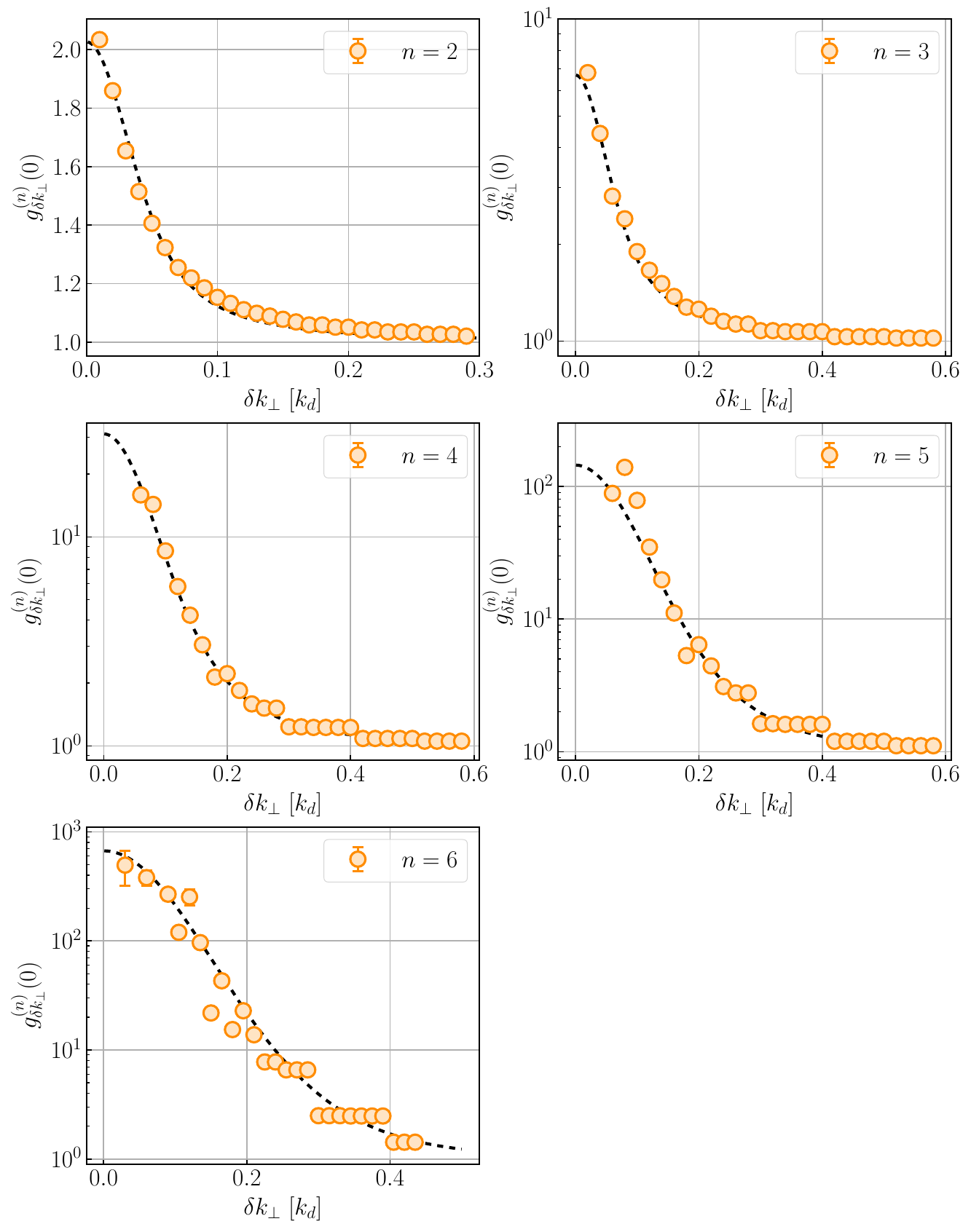}
\caption{Plot of the magnitude $g^{(n)}_{\delta k_{\perp}}(0)$ as a function of the transverse integration $\delta k_{\perp}$. The different panels correspond to different orders $n$ of correlation functions, from $n=2$ to $n=6$. Apart from the data for $n=2$ plotted in linear scale, the vertical axes are in log scale.}
\label{fig-supp1}
\end{figure}

We repeat the above analysis for voxels of varying volume, by changing the transverse size $\delta k_{\perp}$, with the results plotted in Fig.~\ref{fig-supp1}. These results illustrate the fact that the magnitudes of the correlation functions are reduced when the voxel used to compute the statistics is too large, {\it i.e.} of the order or larger than the coherence volume $V_{c}$. 

The fully-contrasted magnitude $g^{(n)}(0)$, expected in voxels $V_{\Omega} \ll V_{c}$, is the value of  $g^{(n)}_{\delta k_{\perp}}(0)$ in the limit $ \delta k_{\perp} \rightarrow 0$. We fit the measured variations of $g^{(n)}_{\delta k_{\perp}}(0)$ with $\delta k_{\perp}$ (see dashed-line in Fig.~\ref{fig-supp1}) and extrapolate $g^{(n)}(0)$ in the limit $ \delta k_{\perp} \rightarrow 0$.

\section{Probability distribution and magnitudes of correlation functions in Bose superfluids}\label{AppC}

The magnitudes of the normalized correlation functions are obtained from the factorial moments of the atom number $N_{\Omega}$ in the small volume $V_{\Omega} \ll V_{c}$. In the case of the BEC mode at ${\bf k}={\bf 0}$, the average number of atoms in $V_{\Omega}$ is relatively large, $\langle N_{\Omega} \rangle \sim 5$. As a consequence, the statistics in $V_{\Omega}$ is sufficient to extract accurately the factorial moments. 

The $n$-factorial moment is defined as $\langle (N_{\Omega})_{n} \rangle=\langle N_{\Omega}(N_{\Omega}-1)...(N_{\Omega}-n+1)\rangle$ and $g^{(n)}(0)$ are expressed as
\begin{equation}
g^{(n)}(0)=\frac{\langle N_{\Omega}(N_{\Omega}-1)...(N_{\Omega}-n+1)\rangle}{\langle N_{\Omega}\rangle^n}=\frac{\langle (N_{\Omega})_{n} \rangle}{\langle N_{\Omega}\rangle^n}
\end{equation}

We define the error bars on $g^{(n)}(0)$ as $\Delta (N_{\Omega})_{n}/\langle N_{\Omega} \rangle^n$, where $\Delta (N_{\Omega})_{n}$ is the standard error of the factorial moment of order $n$, 
\begin{equation}
\Delta (N_{\Omega})_{n}= \frac{1}{\sqrt{N_{\rm runs}}} \sqrt{\langle (N_{\Omega})_{n}^2 \rangle - \langle (N_{\Omega})_{n} \rangle^2} 
\end{equation}

\section{Randomized data and upper limit of the order of correlations}\label{AppD}

We use the randomized data to estimate up to which order $n$ we are able to compute the magnitudes of correlation functions correctly. As explained in the main text, the randomized data sets are built from the actual data sets of the experiment by shuffling the atoms from a given shot into many different shots. By construction, the randomized data has the same number of shots and of atoms per shot than the non-randomized data. Therefore we expect it to provide a means to test the accuracy of the finite data sets we use: because the randomized data should exhibit a perfect Poisson statistics, a systematic deviation to the latter is used as a signal of insufficient statistics.
\\

\begin{figure}[h!]
\includegraphics[width=\columnwidth]{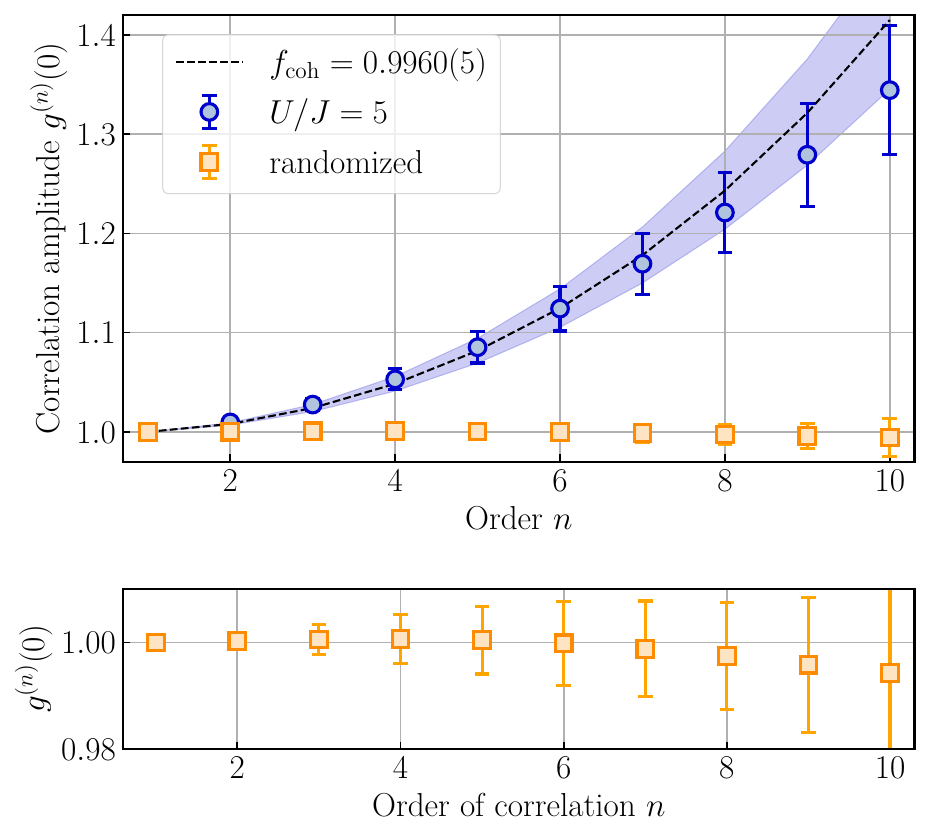}
\caption{Magnitude $g^{(n)}(0)$ as a function of the order $n$ of correlations. The blue circles correspond to experimental data measured in Bose superfluids at $U/J=5$. The orange squares correspond to the randomized data set.}
\label{fig-supp2}
\end{figure}

As illustrated in Fig.~\ref{fig-supp2} in the case of Bose superfluids at $U/J=5$ (same data set as the one shown in Fig.~\ref{fig1} of the main text), we are able to compute correlations of order larger than $n=6$. However, for $n>6$ we observe a systematic shift to the expected value $g^{(n)}(0)=1$ in the randomized data. We attribute this systematic shift to the fact that there are too few shots with an atom number $N_{\Omega} >6$ to correctly evaluate the probability to find $N_{\Omega} =n$ for $n>6$. We therefore restrict our analysis to the orders $n\leq6$. 
A similar analysis for the other data sets shown in this work yields the same result.
	
\section{Model for the magnitudes of correlations in Bose superfluids}\label{AppE}

We introduce a simple model to account for the presence of atoms belonging both to the BEC and to its depletion in the central voxel located at ${\bm k}={\bf 0}$. The depletion of the condensate results from the sum of the quantum depletion (induced by interactions) and the thermal depletion (induce by the finite temperature). As we have no mean to distinguish between the quantum and the thermal depletion, the depletion of the condensate we refer to in the following is the sum of both contributions.

We assume that the statistical properties of the BEC mode are those of a coherent state, $g^{(n)}_{\rm BEC}(0)=1$. In contrast, the statistical properties of the depletion are assumed to be those of a thermal state, $g^{(n)}_{\rm dep}(0)=n!$. This latter assumption holds because we consider a small voxel of size $dk=0.025 k_{d} \ll l_{c} \simeq 0.15 k_{d}$ where the correlation length $l_{c}$ is set by the inverse in-trap size $1/L$ of the gas (see the main text). 

Moreover, we further assume that the BEC and depletion operators are uncorrelated. This is valid because the voxel we consider is much smaller than the volume occupied by the depletion, implying that the correlation between the total number of BEC atoms and the total number of depleted atoms (in the canonical ensemble) can be safely neglected.
\\

We are interested in the statistical properties of the atom number $\hat{N}_{\Omega}= a_{\Omega}^{\dagger} a_{\Omega}$ falling in the considered voxel, where $a_{\Omega}= a_{\rm BEC} +a_{\rm dep}$ is the sum of the operators associated with the BEC and the depletion. Since we assume the operators $a_{\rm BEC}$ and $a_{\rm dep}$ are uncorrelated, one simply obtains
\begin{equation}
\langle (a_{\Omega}^{\dagger})^n a_{\Omega}^n \rangle = \sum_{p=1}^n \binom{n}{p}^2 \langle (a_{\rm BEC}^{\dagger})^p a_{\rm BEC}^p (a_{\rm dep}^{\dagger})^{n-p} a_{\rm dep}^{n-p}\rangle
\end{equation}
where by definition $a_{\Omega}^n=\sum_{p=1}^n \binom{n}{p} a_{\rm BEC}^p a_{\rm dep}^{n-p}$. This leads to the formula given in the main text, 
\begin{eqnarray}
g^{(n)}(0) - 1 &=& \frac{\langle (a_{\Omega}^{\dagger})^n a_{\Omega}^n \rangle }{ \langle a_{\Omega}^{\dagger} a_{\Omega} \rangle ^n} - 1 \nonumber \\
&=& \sum_{p=1}^{n-1} \left [ (n-p)! \binom{n}{p}^2 -\binom{n}{p} \right ] f_{\rm coh}^p (1-f_{\rm coh})^{n-p} \nonumber 
\end{eqnarray}
\\

%
%
%
%
%

\section{Relation $f_{\rm coh}$ versus $f_{c}$.}\label{AppF}

The coherent fraction is approximatively given by
\begin{equation}
f_{\rm coh} \simeq \frac{\rho_{\rm BEC}(0) }{\rho_{\rm BEC}(0) + \rho_{\rm dep}(0)}
\end{equation}

where $\rho_{\rm BEC}(0)$ ({\it resp.} $\rho_{\rm dep}(0)$) is the momentum density of the BEC ({\it resp.} of the depletion) at the center of the Brillouin zone, $\vec{k}=\vec{0}$. The condensate fraction is approximately given by
\begin{equation}
f_c \simeq \frac{\rho_{\rm BEC}(0) \mathcal{V}_{c}}{\rho_{\rm BEC}(0) \mathcal{V}_{c} + \rho_{\rm dep}(0) \mathcal{V}_d }
\end{equation} 

where $\mathcal{V}_{c}$ ({\it resp.} $\mathcal{V}_{d}$) is the total volume of the momentum-space occupied by the BEC ({\it resp.} by the depletion) normalized to the density at $\vec{k}=\vec{0}$.
Combining these two definitions, one obtains the equation Eq.~(\ref{Eq:f_coh}) of the main text.
\newline

To compare the prediction of Eq.~(\ref{Eq:f_coh}) with the measurements, we need to evaluate the volumes $\mathcal{V}_{c}$ and $\mathcal{V}_{d}$. We describe the BEC density profile with an isotropic function \cite{stenger1999}
\begin{equation}
n_{\rm BEC}({\bf k})=\rho_{\rm BEC}(0) e^{-k^2 /2 \sigma_{\rm bec}^2 }
\end{equation}

and the density profile of the depletion with a lorentzian function along each (uncoupled) lattice axis, 
\begin{equation}
n_{\rm dep}({\bf k})= \rho_{\rm dep}(0) \ \Pi_{j=x, y, z} \left ( \frac{ \sigma_{\rm dep}^2/4}{k_{j}^2 + (\sigma_{\rm dep}/2)^2} \right ).  
\end{equation}

We obtain $\mathcal{V}_{c}=(\sqrt{2 \pi} \sigma_{\rm bec})^3$ and $\mathcal{V}_d=\left (\sigma_{\rm dep} \arctan \left [ \frac{k_{d}}{ \sigma_{\rm dep}} \right ] \right )^3$. From fitting the measured density profiles with the above dependency, we obtain $\mathcal{V}_c/\mathcal{V}_d = 0.03(1)$, a value which we use to plot the theoretical prediction in Fig.~\ref{fig4}.

\end{document}